\documentclass{article}
\usepackage{spconf,amsmath,graphicx}
\usepackage{amsmath,amssymb,color}

\definecolor{nblue}{rgb}{0,0,0.6}
\definecolor{nred}{rgb}{0.8,0,0}
\definecolor{ngreen}{rgb}{0,0.4,0}
\definecolor{norange}{rgb}{1.0,0.5,0.3}

\definecolor{scblue}{rgb}{0,0,0.4}

\newcommand{\Bf}[1]{{\bf #1}}

\newcommand{\BS}[1]{\boldsymbol{#1}}

\newcommand{\Brack}[1]{\left[ #1 \right]}

\newcommand{\Trns}[1]{{\bf #1}^{\mbox{\sf \tiny T}}}

\newcommand{\ExP}[3]{#1_{#2}^{(#3)}}

\newcommand{\Trace}[1]{\,\mbox{tr}\!\left\{ #1 \right\}}

\begin{document}


\title{Position dependent prediction combination for intra-frame video coding}

\name{Amir Said \qquad Xin Zhao \qquad Marta Karczewicz \qquad Jianle Chen \qquad Feng Zou}
\address{Qualcomm Technologies, Inc., San Diego, CA, USA}

\maketitle

\begin{abstract}
Intra-frame prediction in the High Efficiency Video Coding (HEVC) standard can be empirically improved by applying sets of recursive two-dimensional filters to the predicted values. However, this approach does not allow (or complicates significantly) the parallel computation of pixel predictions. In this work we analyze why the recursive filters are effective, and use the results to derive sets of non-recursive predictors that have superior performance. We present an extension to HEVC intra prediction that combines values predicted using  non-filtered and filtered (smoothed) reference samples, depending on the prediction mode, and block size. Simulations using the HEVC common test conditions show that a 2.0\% bit rate average reduction can be achieved compared to HEVC, for All Intra (AI) configurations.
\end{abstract}

\begin{keywords}
HEVC, video coding, linear prediction, intra-frame coding
\end{keywords}

\section{Introduction}

Intra-frame coding in the HEVC video coding standard employs the conventional combination of predictive and transform coding, where prediction is based on single reference pixel lines, from previously reconstructed blocks~\cite{Sullivan:12:hev,Lainema:12:ich}. To attain improved compression efficiency while maintaining low decoder complexity, only very simple linear predictors are used, but the encoder can choose among two non-directional and 33~directional predictors.


This approach has proved to be quite effective, but one of its limitations is that the same form of prediction is used for blocks sizes that can range from $4\times 4$ to $64\times 64$, and while exact directional prediction can work well in the small blocks, it is much less effective in larger blocks.

In HEVC this problem is partially addressed by applying a smoothing filter to the reference lines, with a table-based dependence on predictor type and block size~\cite{Wien:15:HEV}. However, this approach does not address the issue that, in reality, less smoothing is needed for pixels near the top-left block borders, and more for pixels further away from the borders. Furthermore, even for blocks of the same size, there is much statistical variability, and additional coding gains can be obtained by allowing the encoder to test more filter choices.

One interesting solution is to apply sets of recursive filters to smooth the predicted values~\cite{McCann:10:vct,Han:10:ivc,Chen:13:are,Li:14:rdo}, since this can naturally provide, up to a certain degree, the desired smoothing control, and experimental results demonstrate that it can improve compression efficiency significantly. 

The computational complexity of the recursive filters can appear very small, if we measure it with conventional techniques. However, in custom hardware and general purpose processors, video applications are now employing parallel computation. Thus, it is advantageous to use techniques that allow a large number of block pixels to be predicted in parallel, simultaneously. Under this consideration, the recursive filtering approach becomes much less attractive.

We propose an extension to HEVC intra prediction, which was accepted to the latest HM-KTA software,\footnote{https://vceg.hhi.fraunhofer.de/svn/svn\_HMKTASoftware/tags/ HM-14.0-KTA-2.0/} called Position-Dependent Prediction Combination (PDPC). It is meant to address all those issues---while maintaining low complexity and efficient parallel implementation---by employing a simple weighted combination of unfiltered and filtered (smoothed) references, where the weights depend on pixel position, prediction mode, and block size.

In the next sections we show how statistical analysis, plus visualization, can help identify what is the type and degree of smoothing needed in the different cases.

\section{Exploiting geometric and statistical features}

\begin{figure}
\centering
\includegraphics[scale=1.0]{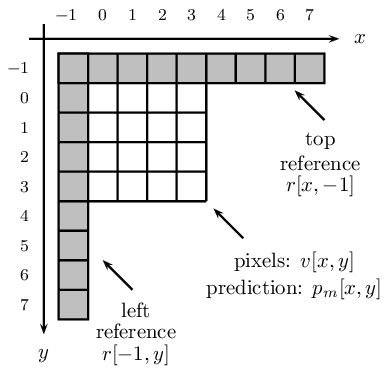}
\caption{\label{fg:HEVCblock}Configuration used for intra-frame prediction in the HEVC video coding standard.}
\end{figure}

Fig.~\ref{fg:HEVCblock} shows the HEVC configuration for intra-frame prediction, for $4\times 4$ blocks, and the notation we use. In an $N\times N$ block, we have the pixels values represented by $v[x,y]$, and their set of predicted values by $p_m[x,y]$ with $x,y\in\{0,1,\ldots,N-1\}$, and $m\in\{0,1,\ldots,34\}$ is the index identifying the predictor type (planar, DC, directions 0 to 33). The reference pixels, which are used for prediction, are represented by a two-dimensional array $r[x,y]$, but with the assumption that only values $r[x,-1]$ and $r[-1,y]$, with $x,y\in\{-1,0,\ldots,2N-1\}$ are used~\cite{Wien:15:HEV}.

The predictor formulas used by HEVC are designed to exploit geometric features in the blocks. Theoretically, we can improve performance using statistics to design better predictors, but by itself this approach tends to produce solutions that are either not effective, or too complex for the demanding video applications. 

We show next that a hybrid approach, exploiting statistical distributions conditioned to geometrical features, can be quite powerful, and even if it cannot be directly used, it yields valuable information on how to design better low complexity predictors.

We can employ the HEVC encoder as a classifier of geometric features, and analyze the statistics of the blocks where a predictor with index $m$ was chosen. To simplify the notation, we represent the block pixel, predicted, and reference values as columns vectors $\Bf{v}$, $\Bf{p}_m,$ and $\Bf{r}$, with dimensions $N^2$, $N^2$, and $4N+1$, respectively. Using this notation, we can use training to estimate the matrices
\begin{equation}
 \label{eq:CorrDefn}
 \Bf{P}_m = E_{\Bf{r}|m}\{ \Bf{r} \, \Trns{r} \}, \quad \Bf{Q}_m = E_{\Bf{v},\Bf{r}|m} \{ \Bf{v} \, \Trns{r} \},
\end{equation}
where operators $E_{\Bf{r}|m}$ and $E_{\Bf{v},\Bf{r}|m}$ define expectation on $\Bf{v}$ and $\Bf{r}$, conditioned to $m$ being chosen, by the HEVC encoder, as best predictor index for those block pixel and reference values.

With those statistics we can compute the linear least-squared-error predictor as
\begin{equation}
  \Bf{p}_m = \Bf{H}_m \Bf{r}, \quad \Bf{H}_m = \Bf{Q}_m \Bf{P}_m^{-1}.
\end{equation}

Using matrix $\Bf{H}_m$ for practical prediction is not feasible because it has $N^2(4N+1)$ elements, and we need one matrix for each predictor index $m$, and for each block size. However, unlike matrix $\Bf{Q}_m$, which provides only the level of similarity between pixels and reference values, matrix $\Bf{H}_m$ can show which similarities are the best to exploit for prediction, by identifying the elements with largest magnitudes in the matrix.

\begin{figure}
\centering
\includegraphics[width=84mm]{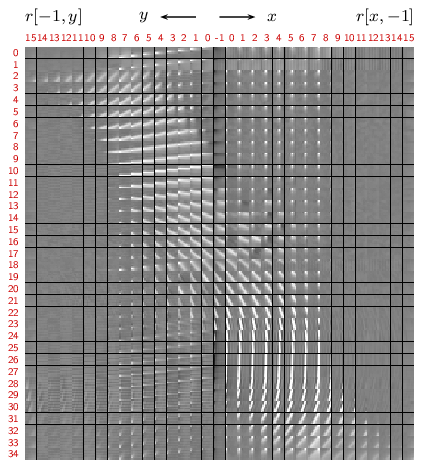}
\caption{\label{fg:MatArray}Example of an image for visualizing the distribution of coefficients in predictor matrices $\Bf{H}_m$. Zeros are represented as gray, positive as lighter, and negative as darker pixels.}
\end{figure}

This can be observed by using a simple technique to visualize the properties of the optimal predictors in matrices $\Bf{H}_m$, and how they vary with $m$. Fig.~\ref{fg:MatArray} shows an example of such $\Bf{H}_m$ visualization image, where gray level represents the weight of a reference pixel (block columns), for a given predictor with index $m$ (block rows). Each $8 \times 8$ block represents a column of $\Bf{H}_m$, but reorganized to represent the corresponding pixel position.

If we create a similar image, with a predictor matrix $\Bf{H}$ equal to the HEVC predictors, we would see that for $m=2,\ldots,34$, we would have well-defined straight lines, according to the direction of those predictors. In Fig.~\ref{fg:MatArray} we can observe the same directional patterns (it was, after all, obtained using HEVC classification for training), but we can also observe that the lines have different degrees of ``blurring,'' to compensate for the fact that in real video frames, patterns commonly deviate from perfectly straight lines.

From this observation, the advantage of applying recursive filters to the predicted values become quite clear---it makes the result of HEVC predictions more similar to the optimal (but highly complex) linear predictors in matrices $\Bf{H}_m$.

Going further, we can observe that most of the coefficients in $\Bf{H}_m$ are very small, and the important coefficients occur with predictable patterns. In the next section we exploit this fact to design new practical predictors.

\section{Position-Dependent Prediction Combination (PDPC)}\label{sc:pdpc}

There are many ways to approximate the features of the predictors in matrices $\Bf{H}_m$. In our approach we considered only the solutions that
\begin{itemize}
\item Combine statistical analysis with the geometry-based predictors in HEVC to approximate the features of optimal predictors in matrices $\Bf{H}_m$;
\item Avoid radical changes in the prediction methodology, and if possible employ the methods already develped for HEVC;
\item Have a formulation that allows parallel computation of all predicted pixels, for all block sizes.
\end{itemize}

Due to the last requirement, we cannot use recursive filters to vary the support of reference pixels that are used to compute each pixel prediction. Nevertheless, we can obtain similar results if we use more than one set of filtered references, and combine them according to the position of the predicted pixel in a block.

This technique, which we call Position-Dependent Prediction Combination (PDPC), can be used for any number of combinations, but we consider having only the original $\Bf{r}$, and one additional filtered version of $\Bf{r}$, which we represent by $\Bf{s}$. Given any two set of pixel predictors $p_{\Bf{r}}[x,y]$ and $q_{\Bf{s}}[x,y]$, computed using only the unfiltered and filtered references $\Bf{r}$ and $\Bf{s}$, respectively, the combined predicted value of a pixel, denoted by $p[x,y]$ , is defined by (we omit the dependence on $m$ to simplify notation)
\begin{equation}
  p[x,y] = c[x,y] p_{\Bf{r}}[x,y] + (1 - c[x,y]) q_{\Bf{s}}[x,y],
\end{equation}
where $c[x,y]$ is a set of weights, computed to approximate predictors in a matrix $\Bf{H}$.

This general approach can be easily adapted for practical requirements. For example, we can use the predictors already defined by HEVC, denoted by $\ExP{p}{\Bf{x}}{\mbox{\scriptsize HEVC}}$, to replace  $p_{\Bf{r}}[x,y]$ and $q_{\Bf{s}}[x,y]$,
and we can define weights using simple formulas.

One additional prediction feature, clearly seen in predictor visualization images as in Fig.~\ref{fg:MatArray}, is that, even for the directional modes, there is a strong correlation between a pixel value and its nearest pixel in the reference set, which is seen to decay in more or less exponential manner.

Putting these two properties together, one choice that produces a good approximations of features in trained predictors $\Bf{H}$, reuses HEVC methods, and is easy to compute, is
\begin{eqnarray}
 p[x,y] & = & \frac{\ExP{c}{1}{v} r[x,-1] - \ExP{c}{2}{v} r[-1,-1] } { 2^{y/d_v} } + \nonumber \\
        & + & \frac{\ExP{c}{1}{h} r[-1,y] - \ExP{c}{2}{h} r[-1,-1] } { 2^{x/d_h} } + \nonumber \\
        \label{eq:PDPCa}
        & + & t[x,y] \; \ExP{p}{\Bf{r}}{\mbox{\scriptsize HEVC}} + b[x,y] \; \ExP{p}{\Bf{s}}{\mbox{\scriptsize HEVC}}
\end{eqnarray}
where
\begin{equation}
 b[x,y] = 1 - \frac{\ExP{c}{1}{v} - \ExP{c}{2}{v} } { 2^{y/d_v} }
            - \frac{\ExP{c}{1}{h} - \ExP{c}{2}{h} } { 2^{x/d_h} } - t[x,y].
\end{equation}
is a normalization factor, and
\begin{equation}
 t[x,y] = \frac{N - \min(x,y)}{N}.
\end{equation}

The first terms in eq.~(\ref{eq:PDPCa}) are meant to exploit correlations at the block boundaries, similarly to recursive filters, but the fast exponential decay of its weights allow for more flexibility in its combination with other predictor term.


The parameters $\ExP{c}{1}{v}$, $\ExP{c}{2}{v}$, $\ExP{c}{1}{h}$, $\ExP{c}{2}{h}$, $d_h$ and $d_v$ can be determined via training (more details shortly). To reduce the number of parameters we also used the rule
\begin{equation}
 d_v = d_h = \left\{ \begin{array}{ll}
  1, & \mbox{if 16 $\times$ 16 or smaller} \\
  2, & \mbox{if 32 $\times$ 32 or larger}
  \end{array} \right.
\end{equation}

Note that when we employ only linear predictors (like HEVC's), we can avoid having to compute both $\ExP{p}{\Bf{r}}{\mbox{\scriptsize HEVC}}$ and $\ExP{p}{\Bf{s}}{\mbox{\scriptsize HEVC}}$, and use
\begin{eqnarray}
 p[x,y] & = & \frac{\ExP{c}{1}{v} r[x,-1] - \ExP{c}{2}{v} r[-1,-1] } { 2^{y/d_v} } + \nonumber \\
        & + & \frac{\ExP{c}{1}{h} r[-1,y] - \ExP{c}{2}{h} r[-1,-1] } { 2^{x/d_h} } + \nonumber \\
        & + & b'[x,y] \ExP{p}{\Bf{s}}{\mbox{\scriptsize HEVC}}
\end{eqnarray}
where
\begin{equation}
 b'[x,y] = 1 - \frac{\ExP{c}{1}{v} - \ExP{c}{2}{v} } { 2^{y/d_v} } -
          \frac{\ExP{c}{1}{h} - \ExP{c}{2}{h} } { 2^{x/d_h} }.
\end{equation}
and
\begin{equation}
 \Bf{s} = a \, \Bf{r} + (1 -a) (\Bf{h}_k \ast \Bf{r}).
\end{equation}
where $a$ is another parameter obtained from training and $\Bf{h}_k$ is a selected filter parameterized by index $k$. In our experimental tests we used this last formulation with binomial filters~\cite{Aubury:95:bnf} for $\Bf{h}_k$, where $k$ is the order of the filter, also optimized from training.

The optimization corresponds to finding the PDPC parameter vector $\BS{\alpha}$ that, given the data classification defined by the HEVC encoder, minimizes the residual mean squared error. Using this approach we can exploit the fact that, since PDPC is linear, it has a corresponding prediction matrix $\tilde{\Bf{H}}_m(\BS{\alpha})$, and the squared error minimization corresponds to
\begin{equation}
 \min_{\BS{\alpha}} \Brack{ \Trace{\tilde{\Bf{H}}_m(\BS{\alpha}) \Bf{P}_m \tilde{\Bf{H}}^{\mbox{\sf \tiny T}}_m(\BS{\alpha})  } - 2 \Trace{ \tilde{\Bf{H}}_m(\BS{\alpha}) \Trns{Q}_m } },
\end{equation}
where $\Bf{P}_m$ and $\Bf{Q}_m$ are defined in eq.~(\ref{eq:CorrDefn}).

\begin{figure}
\centering
\includegraphics[width=84mm]{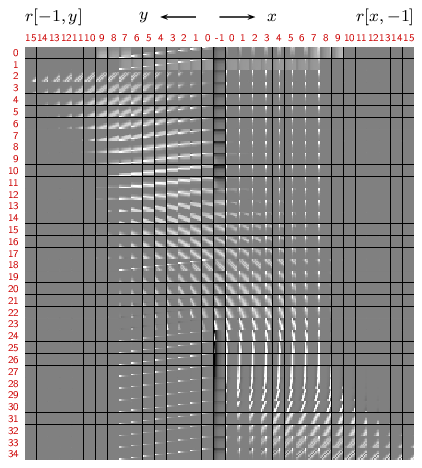}
\caption{\label{fg:PdpcArray}Image for visualizing the PDPC predictor matrices ($\tilde{\Bf{H}}_m(\BS{\alpha})$), where we can observe how they approximate the matrices in Fig.~\ref{fg:MatArray}, but with a much smaller number of parameters.}
\end{figure}

\section{Experimental results}\label{sc:result}

\begin{table}
\centering
\caption{\label{tb:FourC}PDPC coding gains over HEVC, version with four sets of predictor parameters.\vspace{2mm}}

\begin{tabular}{|l|c|c|c|} \hline
 & Y & U & V \\ \hline \hline
 Class A & --2.24\% & --1.80\% & --1.83\% \\ \hline
 Class B & --2.26\% & --1.74\% & --1.90\% \\ \hline
 Class C & --1.59\% & --1.45\% & --1.66\% \\ \hline
 Class D & --1.54\% & --1.29\% & --1.38\% \\ \hline
 Class E & --2.48\% & --3.65\% & --3.44\% \\ \hline
 \bf Average & \bf --2.01\% & \bf --1.89\% & \bf --1.97\% \\ \hline
\end{tabular}
\end{table}

\begin{table}
\centering
\caption{\label{tb:FourHDR}PDPC coding gains over HEVC, version with four sets of predictor parameters, UHD $3840 \times 2160$ sequences.\vspace{2mm}}

\begin{tabular}{|l|c|c|c|} \hline
 UHD Sequence & Y & U & V \\ \hline \hline
 EBU LupoBoa      & --4.56\% & --7.68\% & --7.89\% \\ \hline
 EBU StudioDancer & --4.18\% & --6.32\% & --6.00\% \\ \hline
 EBU VeggieFruits & --3.45\% & --5.75\% & --5.07\% \\ \hline
 \bf Average      & \bf --4.06\% & \bf --6.58\% & \bf --6.32\% \\ \hline
\end{tabular}
\end{table}

\begin{table}
\centering
\caption{\label{tb:TwoC}PDPC coding gains over HEVC, version with two sets of predictor parameters.\vspace{2mm}}

\begin{tabular}{|l|c|c|c|} \hline
 & Y & U & V \\ \hline \hline
 Class A & --1.54\% & --1.20\% & --1.28\% \\ \hline
 Class B & --1.53\% & --1.25\% & --1.28\% \\ \hline
 Class C & --1.29\% & --1.00\% & --1.05\% \\ \hline
 Class D & --1.10\% & --0.89\% & --0.88\% \\ \hline
 Class E & --1.85\% & --2.69\% & --2.54\% \\ \hline
 \bf Average & \bf --1.45\% & \bf --1.34\% & \bf --1.34\% \\ \hline
\end{tabular}
\end{table}

Using the training procedure described above, we obtained sets of PDPC parameters that approximate the optimal prediction matrices $\Bf{H}_m$. This can be observed by applying the same visualization technique of Fig.~\ref{fg:MatArray} to the PDPC prediction matrices $\tilde{\Bf{H}}_m(\BS{\alpha})$, an shown in the example of Fig.~\ref{fg:PdpcArray}.

The proposed prediction method has been integrated into the latest HEVC reference software (HM~16.6),\footnote{https://hevc.hhi.fraunhofer.de/svn/svn\_HEVCSoftware /tags/HM-16.6/} and tested using All Intra (AI) coding configuration and main 10 profile, as specified in the the common tests conditions defined during development HEVC, and evaluated using the Bj{\o}ntegaard-Delta (BD) coding gain measure~\cite{Bossen:11:ctc,Bjontegaard:01:cap}. We tested two versions, which employ two and four sets of predictor parameters, tested by the encoder, and the selected optimal one is signaled at Coding Unit (CU) level. In both cases, one of the options is to use the HEVC predictors directly (i.e., with PDPC). A slightly more optimized version is available for testing in the HM-KTA software.

Table~\ref{tb:FourC} shows the results for the version with four sets of predictor parameters. We can observe that PDPC yields an average coding gain around 2\%. Even larger gains (around 4\%) can be obtained on UHD sequences, as shown in Table~\ref{tb:FourHDR}. Table~\ref{tb:TwoC} shows that the version with two sets of parameters yield smaller, but still significant, gains.


\end{document}